\documentclass{article} 
\begin{document} 
\begin{center} 
\title{Molecular quantum mechanical observers, symmetry, and string theory} 
\author{M.Dance}
\maketitle 
\end{center} 

\begin{abstract} 
The paper~\cite{Dance0601} tentatively suggested a physical picture that might underlie string theories. The string parameters $\tau $ and $\sigma_i $ were interpreted as spacetime dimensions which a simple quantum mechanical observer can observe, while symmetries of the relevant observer states could limit the observability of other dimensions.  An atomic observer was the focus of the discussion. The present paper extends the discussion of\cite{Dance0601} to molecular observers, including the nature of some common molecular bonds and their symmetries. 
\end{abstract} 

\section{Introduction}

Heisenberg noted that his uncertainty principle applies to observers as well as to observed systems~\cite{heisenberg}. He postulated that any practical effect of observer indeterminacy could be eliminated by allowing the observer's mass to approach infinity, but gravity would then become important.  To arrive at a theory that unifies quantum mechanics and gravity, it might therefore be necessary to include quantum mechanical observers.  

Physically, non-commutative geometries might incorporate quantum mechanical uncertainty in the position and momentum of an observer¡¯s centre of mass.  For example, Bander~\cite{bander} has shown that non-commutative geometry is an effective low-energy theory of systems coupled to an auxiliary system; one might interpret the auxiliary system of~\cite{bander} as an observer.  
 
It then seems possible that a richer theory remains to be uncovered, one which includes properties of a quantum mechanical observer¡¯s states. In~\cite{Dance0601}, it was suggested that a quantum mechanical observer term in a Lagrangian density corresponds to a string theory if symmetries of relevant observer states make corresponding spacetime dimensions unobservable to the observer, and if the radial dimension $r$ is not observable by such an observer, requiring instead an intelligent observer. These ideas were discussed in particular for atomic quantum mechanical observers.  The paper~\cite{Dance0601} tentatively suggested that this physical picture might even underlie string theories. A recent review of some other work on quantum reference frames is at~\cite{Bartlett0610}.

The present paper extends the discussion of~\cite{Dance0601} to quantum mechanical observers that observe using one or more molecular bonds.

\section{Summary of the physical picture}

This section briefly summarises~\cite{Dance0601}.  To incorporate observer dynamics, a simple observer term was added in~\cite{Dance0601} to a field theory Lagrangian density in a flat classical background spacetime. The term represents the dynamics of the centre of mass of the quantum mechanical observer O1:-

\begin{equation}
L^{EK}_{obs} = \frac{1}{2}m \eta_{ij}\frac{dX^i}{dT}\frac{dX^j}{dT}
\end{equation}
which corresponds to a Lagrangian density
\begin{equation}
\mathcal{L} =  \frac{1}{2}m \kappa^{\alpha \beta } \eta_{\mu \nu } 
\frac{\partial{X^\mu}}{\partial{x^\alpha}}  
\frac{\partial{X^\nu}}{\partial{x^\beta}}
\label{L}
\end{equation}
where $m$ is the mass of the quantum observer O1, $x^{\alpha }$ and $x^{\beta }$ are the coordinates ($t$, $x$, $y$, $z$) or e.g. ($t$, $r$, $\theta$, $\phi$) internal to the observer, and ${X^{\mu }}$ are fields representing the observer's centre of mass position in the reference system of another observer O2 that is observing the first, quantum mechanical observer.  The factor $\kappa $ is given by: 
\begin{equation}
\kappa ^{\alpha \beta} = \rho(x) \frac{dx^\alpha}{dT_i} \frac{dx^\beta}{dT_i}
\end{equation}
where $\rho(x)$ is a function of $x$.

The simplest string action is the area of the world-sheet of a string propagating through flat space-time, multiplied by a constant tension:-
\begin{equation}
S_{string} = -\frac{T}{2} \int d\tau d\sigma
\eta^{\alpha \beta } \eta_{\mu \nu } 
\frac{\partial{X^\mu}}{\partial{x^\alpha}}
\frac{\partial{X^\nu}}{\partial{x^\beta}},
\label{str}
\end{equation}
where $X^{\mu }$ are the space-time coordinates of a point on the string, and
$x^{\alpha }$, $x^{\beta}$ are members of ${\tau, \sigma}$.

Focusing on the derivatives in Equations~(\ref{L}) and~(\ref{str}) above, string theory uses the worldsheet parameters ($\tau$, $\sigma_i$ ) instead of internal observer space-time coordinates.  In \cite{Dance0601}, it was postulated that a physical correspondence can be made between the internal O1 coordinates ($t$, $r$, $\theta$, $\phi$) and ($\tau$, $\sigma_i$), as follows.

It was postulated in~\cite{Dance0601} that an observer will not be able to extract information about parameters with respect to which the first, quantum mechanical observer's initial and final states are symmetric in its own internal coordinate system. At the quantum level, a common symmetry is rotational symmetry. An example is an H atom whose initial and final states both possess a spherically symmetric $1s$ electronic state. To this H atom observer, the angular coordinates $\theta $ and $\phi $ may be meaningless.  If the H atom begins and ends an observation instead with its electron in (say) an excited state, such as a $2p_z$ state, then the H atom has symmetry about an internal $z$ axis; it was postulated that the concept of the polar coordinate $\phi $ may be meaningless to this observer, and that this coordinate does not appear in the Lagrangian density.  The more symmetries an observer possesses, the less the observer may be able to detect. It was also postulated that a simple quantum mechanical observer such as an atom cannot measure the radial coordinate $r$.

In this way, a correspondence was made between the internal observer coordinates $(t, r, \theta , \phi)$ and the string-theory parameters $(\tau , \sigma_i)$, where $\tau = t$, and the $\sigma_i$ of string theory correspond to the angular coordinates that the quantum observer can detect.  It was suggested that there are 1D-observers (that can detect one angular coordinate) and 2D-observers (which can detect two angular coordinates).  The 1D-observers would correspond to strings, and 2D-observers to 2-branes. 

The paper~\cite{Dance0901} included the internal dynamics of a fermion field in the quantum mechanical observer, in addition to the centre of mass term above.  It was suggested there that quantum mechanical uncertainties in the transformation between the reference systems of O1 and O2 might require the use of $d$ spinor fields for the fermion term, where $d$ is the number of spacetime dimensions.  With this fermion term, $\mathcal{L}$ becomes:
\begin{equation}
\mathcal{L}  =  \frac{1}{2}m \kappa^{\alpha \beta } \eta_{\mu \nu } 
                \frac{\partial{X^\mu}}{\partial{x^\alpha}}  
                \frac{\partial{X^\nu}}{\partial{x^\beta}} 
                + i\bar{\psi}^\mu \gamma ^{\alpha} \partial_{\alpha} \psi_\mu
\label{EqL} 
\end{equation}
where as before, the $x^{\alpha}$, $x^{\beta}$ are coordinates in O1's reference system.

The above has summarised the basic physical picture described in previous work including~\cite{Dance0601}. The sections below will now extend the discussion to a molecular observer.

\section{Molecular observer - coordinates}

Observations often take place by a change within a molecule, often changing the state of an electron in a molecular bond between two covalently bonded atoms. In the discussion above, the Lagrangian density $\mathcal{L}$ in equation~(\ref{L})  represented the dynamics of the centre of mass of the quantum mechanical observer, which was discussed for an atom in ~\cite{Dance0601}.  For a quantum observer that is a molecule, the relevant coordinates that will be used in $\mathcal{L}$ will be a little different.  We discuss these coordinates in this section.  We will focus the discussion on the parts of the molecule that are most relevant for the observation of the external world.  Typically this will be one or two molecular bonds in the molecule.  

As in~\cite{Dance0601}, the ${x^\alpha}$ will be taken to represent coordinates in the quantum observer O1's internal coordinate system. At any time $x^0$ in O1's reference system, we will take the classical position of the spatial origin of the ${x^\alpha}$ coordinates to be, at the point that gives the relevant (observing) part of the molecule the maximum symmetry.  For example, if the relevant observing part of the O1 is a carbon-carbon covalent bond (e.g. as in the retinal molecule, which is the first observation stage in human vision processes), the origin of the internal observer coordinate system will be taken to be at the centre of that bond.

The ${X^\mu}$ will now represent the coordinates of the centre of mass of each relevant entity in the observer molecule, in the second observer O2's reference system.  Alternatively, the ${X^\mu}$ might represent coordinate differences between relevant entities.  In the example of a carbon-carbon bond, the ${X^\mu}$ could be coordinates of the two nuclei and the two electrons (classically) in O2's reference system.

The Lagrangian density term for the observer O1 is now:
\begin{equation}
\mathcal{L} = \lambda^{\alpha \beta } \eta_{\mu \nu } 
\frac{\partial{X^\mu}}{\partial{x^\alpha}}  
\frac{\partial{X^\nu}}{\partial{x^\beta}}
\label{Lmol}  
\end{equation}
and
\begin{equation}
\lambda ^{\alpha \beta} = \frac{1}{2}\Sigma_i m_i \kappa _i^{\alpha \beta}   
\end{equation}
where $m_i$ is the mass of the $i$th observer component, $x^{\alpha }$ are the coordinates e.g. $(t, r, \theta, \phi)$ internal to the observer O1, and each $\kappa _i$ is given by:
\begin{equation}
\kappa _i^{\alpha \beta} = \rho_i(x) \frac{dx^\alpha}{dT_i} \frac{dx^\beta}{dT_i}
\end{equation}
where each $\rho_i(x)$ is a function of $x$.  
Normally all of the $T_i$ can be taken to be the same time, which can be called $T$, in O2's coordinate system. 

In~\cite{Dance0601} where the quantum observer was an atom, the number of components $X^\mu$ was simply 4.  By now including terms to represent a number of atoms or other entities in the first stage observer molecule, the number of components $X^\mu$ is increased. In string theory, a large number of ${X^\mu}$ components appearing in $\mathcal{L}$ is interpreted as representing a large number of spacetime dimensions; the dimensionality then must be reduced to 4 by compactification.  In the present picture, the interpretation of ${X^\mu}$ is quite different, and there is no need for compactification.

In string theories, particular dimensions of interest are 10 and 26. Let us see how these numbers can be arrived at in the physical picture discussed in the present paper. 
In this physical picture, the set ${X^\mu}$ can have 10 members if we consider three relevant entities (or differences) in the molecular observer, each with the same time coordinate in the second observer's reference system. Each of the three entities has three spatial coordinates, so the total number of coordinates ${X^\mu}$ is $(1 + 3\rm{x}3 = 10)$.  (Of course, the Lagrangian density above is very simple and is not a superstring Lagrangian density.)

The set ${X^\mu}$ can have 26 members if we consider a first stage observer that is comprised of two parts. The first part is as described in the paragraph above, and contributes 10 coordinates as described there. The second part contributes 16 coordinates: one time coordinate and five sets of spatial coordinates, from five entities (or five coordinate differences).  These two parts could perhaps themselves be two sequential quantum mechanical stages of observation.
 
Having described the coordinates, we will next discuss molecular bonds and their symmetries. We will then be in a position to discuss how a molecular observer fits into the framework of~\cite{Dance0601}.
 
\section{Molecular bonds}

We will now discuss molecular bonds and their symmetries, and how rotational symmetries in molecular bonds can fit into the framework in~\cite{Dance0601}.  We first give a simple description of molecular bonds.

Quantum chemistry describes the states of electrons within molecules with varying degrees of complexity. A useful and simple approach is valence bond theory.  Valence bond theory uses the concept of sigma, pi and delta molecular bonds to describe covalent bonds between two adjacent bonded atoms. These sigma, pi, and delta molecular bonds correspond to superpositions of atomic orbitals of the contributing atomic states, with some changes due to the interactions between the atoms.  Texts such as~\cite{Shaik} and~\cite{Webster} contain descriptions of the sigma, pi, and delta bonds. Briefly, the bonds are as follows:
\begin{itemize}
\item{In a sigma bond, the two contributing atomic orbitals overlap and interact end-to-end. For example in the hydrogen molecule, the 1s electrons of the two H atoms form a sigma bond.}
\item{In a pi bond, two lobes of one electron orbital overlap two lobes of the other electron orbital. The two atomic orbitals overlap and interact side-by-side.} 
\item{A delta bond occurs where four lobes of one electron orbital overlap four lobes of the other electron orbital.}
\item{In valence bond theory, single covalent bonds consist of one sigma bond, double bonds consist of one sigma bond and one pi bond, and triple bonds have a sigma bond and two pi bonds.}
\end{itemize}

In all cases, the symmetry of the molecular bond is the same as that of the constituent orbitals, relative to the bond axis.  Other contributing atomic orbital symmetries may be lost.

\section{Molecular observers, symmetry, and string theory}

We now suggest that the postulates in~\cite{Dance0601} can be applied to an observer that is a molecule, whenever the relevant molecular bond has a rotational symmetry. The basic postulate remains the same - that an observer will not be able to extract information about parameters or dimensions with respect to which the observer's initial and final states are symmetric in its own internal coordinate system. Let us call the bond axis the $z$ axis. If an observation occurs by means of a molecular bond, and the relevant electron states in the bond are rotationally symmetric about the $z$ axis, the postulate means that the angle $\phi$ is essentially invisible to the molecular observer.  A $\sigma$ bond has this property, just as a $p_z$ atomic orbital does.

Therefore, it is suggested that a $\sigma$ bond can operate like a $p_z$ orbital, and give rise (in this picture) to a 1D-observer.

The $\pi$ bond has some rotational symmetry. If we postulate that the simple observer will not notice a sign change in the wavefunction, the pi bond has a discrete symmetry under 180 degree rotations.  The complete rotational symmetry about an axis is lost.  In the picture of~\cite{Dance0601}, the $\pi$ bond might therefore give rise to an apparent 2D-observer which can perceive apparent observed membranes that have some kind of constraint in the $\phi$ (i.e. $\sigma_2$) angular dimension.  

We will not discuss delta bonds, because the sigma and pi bonds are by far the most common molecular bonds.

Therefore, for quantum mechanical observers that are molecules, the same suggestions apply here as in~\cite{Dance0601}. This is because many molecular bonds have rotational symmetries that could provide the same effective dimensional reduction as an atomic orbital does.  We obtain similar correspondences with string theory Lagrangian densities as in~\cite{Dance0601}.

\section{Discussion} 

The paper~\cite{Dance0601} added a simple quantum mechanical observer term to a Lagrangian density, and suggested that internal observer symmetries may effectively eliminate some coordinates, internal to the observer, from observation. It was also suggested that the radial dimension $r$ is not observable by such an observer.  It was suggested that such theories may shed light on string theories, e.g. if the string theory parameters $\tau $ and $\sigma_i$ represent the coordinates which the observer can observe. 

The present paper has extended the discussion to quantum mechanical observers that are molecules. The same suggestions apply here as in that paper, as many molecular bonds also have rotational symmetries that could provide the same effective dimensional reduction. It is also suggested that molecular observers might give rise to string theory Lagrangian densities with apparent spacetime dimensionalities greater than 4, for example 10 or 26, and in a physical picture that has no need for subsequent compactification.  

We have here considered $\mathcal{L}$ for O1 to include only the centre-of-mass dynamics of O1, and to exclude potential terms. This is because our focus is on the motion of O1 as a whole due to its role as an observer; our focus is on the transformations between reference frames.  For our purposes, we have taken the structure of O1 to be given and fixed.  This condition could potentially be relaxed in future.   
 
The ideas in~\cite{Dance0601} could perhaps be extended in future to incorporate further symmetries that occur in molecules, and to incorporate other aspects of the greater complexity of molecules as observers.  The simple picture in~\cite{Dance0601} will become more complicated, but the same underlying postulates can be made, and they will reduce the apparent dimensionality and/or other information available to the molecular observer.  For example, molecular bonds may contain some reflection symmetries, e.g. in the perpendicular plane that bisects the bond.  In that case, the observer O1 might not be able to tell the difference between one direction and the mirror image direction reflected in the plane of symmetry.  It might also be useful in future to extend ${x^\alpha}$ to a larger set, with more internal O1 coordinates arising in the molecule from a greater number of degrees of freedom than for an atomic observer. To achieve this, it might be required to postulate that some effect might prevent the use of a single set of 4 spacetime coordinates for O1's internal coordinates, so that the molecule is to be considered as a patchwork of different coordinate systems for each relevant region of O1.  Such an enlarged ${x^\alpha}$ set might give rise to theories that might contain $p$-branes with $p>2$.   

If in future there is taken to be a patchwork set of coordinates for O1, where local times $t_i$ might be taken to differ between regions of O1, there might be a link with two-time physics.  A similar link might arise if it is taken to be possible for O2 coordinates $T_i$ to differ among each other.
 
It might be useful to investigate how one or more later stages of observation, perhaps an intelligent observation stage subsequent to O2, might best be imposed and described. It may also be appropriate to incorporate decoherence mechanisms.

\section{Conclusion}

The present paper has extended the discussion in~\cite{Dance0601} to quantum mechanical observers that are molecules. The same suggestions apply here as in that paper, as many molecular bonds also have rotational symmetries that could provide the same effective dimensional reduction, giving rise to Lagrangian densities that correspond to string theories. It is also suggested that molecular observers might give rise to string theory Lagrangian densities with apparent spacetime dimensionalities greater than 4, for example 10 or 26, and in a physical picture that has no need for subsequent compactification.  There are a number of other directions in which the work in~\cite{Dance0601} could perhaps be taken forwards in future.


\begin{thebibliography}{9}

\bibitem{heisenberg}W. Heisenberg, Zeitschrift fur Physik, 43, 172-98 (1927). English translation reproduced in ¡°Quantum Theory and Measurement¡±, ed. J. Wheeler and W.H. Zurek,  Princeton University Press, 1983.
 
\bibitem{bander} M. Bander, "Coordinate noncommutativity as low energy effective dynamics", hep-th/0501159.

\bibitem{Dance0601} M. Dance, "Symmetry limitations on quantum mechanical observers, and conjectured link with string theory", hep-th/0601104. 

\bibitem{Bartlett0610} S.D. Bartlett, T. Rudolph, R.W. Spekkens, "Reference frames, superselection rules, and quantum information", Rev. Mod. Phys. 79, 555 (2007), quant-ph/0610030.

\bibitem{Dance0901} M.Dance, "Quantum mechanical observer and superstring/M theory", arXiv:0901.0060. 

\bibitem{Shaik} S.S. Shaik, P.C. Hiberty, "A Chemist's Guide to Valence Bond Theory", New Jersey: Wiley-Interscience (2008).
 
\bibitem{Webster} B.Webster, "Chemical Bonding Theory", Blackwell Scientific Publications (1990).



\end{thebibliography}
\end{document}